\documentclass[aps,prb,twocolumn,groupedaddress,showpacs,superscriptaddress,amssymb,amsmath]{revtex4-1}
\usepackage{graphicx}
\usepackage{tabularx}
\usepackage{color}
\usepackage{amsmath}
\usepackage{comment}
\newcommand{\be}{\begin{equation}}
\newcommand{\ee}{\end{equation}}
\newcommand{\bea}{\begin{eqnarray}}
\newcommand{\eea}{\end{eqnarray}}
\usepackage{dcolumn}
\usepackage{hyperref}
\usepackage{bm}
\usepackage{epsf}
\usepackage{subfigure}
\usepackage{epstopdf}%
\usepackage{amsfonts}

\let\la\langle \let\ra\rangle
\let\bs\boldsymbol

\begin{document}

\title{Full counting statistics and coherences: fluctuation symmetry in heat transport with the Unified quantum master equation}
%


\author{Matthew Gerry}
\affiliation{Department of Physics, University of Toronto, 60 Saint George St., Toronto, Ontario M5S 1A7, Canada}

\author{Dvira Segal}
\affiliation{Department of Physics, University of Toronto, 60 Saint George St., Toronto, Ontario M5S 1A7, Canada}
\affiliation{Chemical Physics Theory Group, Department of Chemistry and Centre for Quantum Information and Quantum Control,
University of Toronto, 80 Saint George St., Toronto, Ontario M5S 3H6, Canada}
\email{dvira.segal@utoronto.ca}
\date{\today}

\begin{abstract}
Recently, a ``Unified" quantum master equation was derived and shown to be 
of the Gorini-Kossakowski-Lindblad-Sudarshan (GKLS) form. 
This equation describes the dynamics of open quantum systems in a manner that forgoes 
the full secular approximation and retains the impact of coherences between eigenstates close in energy. 
We implement full counting statistics with the Unified quantum master equation to investigate the statistics 
of energy currents through open quantum systems with nearly degenerate levels. 
We show that, in general, this equation gives rise to dynamics that satisfy fluctuation symmetry, a sufficient condition for the Second Law of Thermodynamics at the level of average fluxes.
For systems with nearly degenerate energy levels, such that coherences build up,  the Unified equation
is simultaneously {\it thermodynamically consistent} and {\it more accurate} than the fully Secular master equation.
We exemplify our results for a ``V" system facilitating energy transport between two thermal baths 
at different temperatures. We compare the statistics of steady-state heat currents through this system as predicted by the 
Unified equation to those given by the Redfield equation, which is less approximate but, in general, not thermodynamically consistent.
We also compare results to the Secular equation, where coherences are entirely abandoned. 
We find that maintaining coherences between nearly degenerate levels is essential for the properly capturing the current and its cumulants.
On the other hand, the relative fluctuations of the heat current, which embody the thermodynamic uncertainty relation, display inconsequential
dependence on quantum coherences. 
\end{abstract}

\maketitle

\section{Introduction}
Quantum master equations (QMEs) play a key role in the study of open quantum systems, 
describing the time evolution of the reduced density operator of a system 
coupled to thermal baths. 
Assuming weak system-bath coupling, QMEs are typically obtained starting from the unitary evolution of the full system plus baths, 
by making the Born-Markov approximation and tracing over environmental degrees of freedom \cite{breuer-book}. 
The resulting Redfield equation may be used in its own right, or serve as a jumping off point for a variety of 
further approximations leading to quantum master equations with differing features \cite{manzano2020,trushechkin2021,farina2019,becker2022}.

Redfield equations are well known not to be of Gorini-Kossakowski-Lindblad-Sudarshan (GKLS) form \cite{GKS,Lindblad}, 
and thus, in general, fail to preserve the positivity of the reduced density operator 
\cite{Tannor97,EPL14,hartmann2020}. 
The secular approximation, which ignores oscillating terms in the dissipative part of the master equation, 
is often employed to avoid this pitfall \cite{hartmann2020,archak2022}. 
Indeed, when neglecting all oscillating terms,  
one obtains master equations
that have a number of advantageous properties with regards to thermodynamics--- but
 also neglect any impact that eigenstate coherences can have on population evolution, and vice-versa. 
This approximation is therefore invalid for systems with eigenstates close in energy, 
in which case some terms in the Redfield equation oscillate on timescales 
that rival the timescales for the dynamics of interest \cite{trushechkin2021}.

Recently, a new form of quantum master equation, termed the ``Unified" QME (UQME), 
has been developed to describe, with greater accuracy than the fully Secular master equation, 
open quantum systems with nearly degenerate energy levels \cite{trushechkin2021}. 
It does so, in part, by making the secular approximation only with respect to pairs of levels separated by an amount of 
energy substantially greater than that characterizing the system-bath interactions.  
For transitions associated with these pairs, 
the oscillating terms in the Redfield equation oscillate 
fast enough to be 
neglected. 
The impact on the dynamics of relevant coherences--those between levels close in energy--is captured by the UQME. 
This master equation has also been shown to be of GKLS form, and thus preserves the positivity of the 
system's reduced density operator, as desired \cite{trushechkin2021}.

In addition to the preservation of positivity, one desirable property of quantum master equation descriptions  
is {\it thermodynamic consistency}: namely, that the equations give rise to dynamics that satisfy the laws of thermodynamics. 
More specifically, QMEs should satisfy quantum fluctuation theorems. 
These theorems serve as a sufficient condition for the Second Law of Thermodynamics at the level of averages 
\cite{Jarz04,saito2008,Gaspard09,esposito2009,hanggi11}, 
as well as familiar results including the Green-Kubo relation between 
steady-state current fluctuations and conductance in the linear response regime \cite{saito2008}. 

With ongoing advances in quantum thermodynamics, questions over the consistency and accuracy of QMEs are gaining much attention:
To realize quantum thermal machines that build on the interplay between coherent and thermal effects 
(see e.g., proposals  \cite{rahav12,seifert15,huber15,raam16,johal17,misha17,kilgour18,junjie21,ivander2022,serra18,novotny18,novotny19,goold19,landi19,hammam21,latune19,latune20,latune21,tromb21}), 
it is imperative to develop methods that are both thermodynamically consistent, and account for the behavior of quantum coherences.
Some of the examined aspects of QME concern their positivity \cite{gernot08,farina2019,trushechkin2021, davidovic20,lidar20,lee22,alicki22}, the ``right" basis  \cite{novotny02,levy14,local17,luis17,chiara18,naseem18,zambrini19,adam21},
and modifications to the weak coupling QME  \cite{juzar12,juzar13,lidar13,gernot20,wacker21} for improving
 consistency and accuracy,
as well as extensions beyond weak coupling and Markovianity, see e.g., \cite{becker2022, geva03, geva06, PT1,PT2,cao16,hava2018,RCrev,Nick1,Nick2,latune22,tanimura20,andres22}. This paper is focused on verifying that the UQME, which accounts for coherences, fulfills the heat exchange fluctuation theorem.

In the language of full counting statistics, the steady state fluctuation symmetry for entropy production
can be written in terms of the moment generating function, $\mathcal{Z}(\bs\chi,t)$, a function of a counting field $\bs\chi$, whose derivatives evaluated at $\bs\chi=0$ give the higher order moments of the heat currents through the system at all times \cite{esposito2009,gernot-book}. In the long time limit, provided the Hamiltonian has no explicit time-dependence, one may consider also the steady-state scaled cumulant generating function, $\mathcal{G}(\bs\chi)$, whose derivatives are instead the heat current cumulants.
The fluctuation symmetry for {\it steady-state} heat transport, in the absence of particle transport,
states that \cite{Jarz04,gernot-book}, 
\be
    \mathcal{G}(\bs\chi) = \mathcal{G}(-\bs\chi-i\bs\beta),
\label{eq:FS}
\ee
where $\bs\beta$ is a vector of inverse temperatures of the thermal baths with which the system exchanges energy.
Below we refer in short to Eq. (\ref{eq:FS}) as the ``fluctuation symmetry". 
Proving this symmetry for the UQME, 
as a consequence of a more general symmetry of the moment generating function, 
and quantifying its violation (and thus the breakdown of transport relationships)
under the Redfield QME is the main focus of this work. 

The heat exchange fluctuation symmetry has been shown to hold for the fully Secular quantum master equation \cite{gernot-book}, 
including in the case where {\it exact} degeneracies are present in the system. In this case, coherences persist between degenerate levels at steady state, and their effect must be included 
in the dynamics \cite{schaller2016}.  
Recent work has also established general conditions for thermodynamic consistency of quantum master equations \cite{esposito2022}.
In what follows, we will show that the UQME, in addition to being of GKLS form and preserving positivity, satisfies the heat exchange
fluctuation symmetry.

We will do so using analytic arguments, and demonstrate these results via numerical simulations pertaining to the so-called ``V" model, 
a simple model consisting of one lower energy level coupled via interactions with bosonic thermal reservoirs to two excited states, 
as shown in Fig. \ref{fig:V-diagram}. 
The excited states are at substantially higher energy than the ground state, but are close enough in energy to one another that 
coherences between them cannot be ignored. 
In fact, the V system has been shown to exhibit long-lived transient coherences when coupled to a single reservoir 
\cite{tscherbul2014,dodin2016a,dodin2016b,dodin2022}, as well as coherences that persist in nonequilibrium steady states 
\cite{li2015,wang2018,wang2019,koyu2021,kilgour18,junjie21,ivander2022}. 
Coherences in this model have a significant effect on the population dynamics and the steady-state transport behavior 
as governed by quantum master equations \cite{kilgour18,junjie21,ivander2022}

Thus, the UQME is a good candidate for describing, accurately and in a {\it thermodynamically consistent manner}, 
the dynamics and steady-state behavior of {\it level populations and coherences} in the V system and similar models. 
It is also effective in describing the statistics of heat currents such systems may facilitate when coupled to multiple baths 
at different temperatures \cite{ivander2022}.

This paper is organized as follows. 
In Sec. \ref{sec:UQME} we review the Unified QME and discuss the conditions for its validity. 
In Sec. \ref{sec:FCS} we implement full counting statistics for the UQME by deriving a version of the master equation that 
is ``dressed" with a counting field. From this equation we are able to derive the cumulant generating function (CGF) 
for the statistics of heat currents. We go on to prove, based on a symmetry of the counting field-dependent generator 
of time translations, that the CGF obtained with the UQME satisfies the heat exchange fluctuation symmetry. 
In Sec. \ref{sec:V_model} we focus on the specific example of a V system facilitating heat transport between two bosonic 
reservoirs at different temperatures, demonstrating that fluctuation symmetry is satisfied when the 
dynamics are modelled by the UQME, and violated when the Redfield equation is used instead. 
We go on to compare the predictions the two equations make for the mean and variance of the heat current 
at steady state by examining
transport symmetries including the Green-Kubo relation, 
as well as the Thermodynamic Uncertainty Relation (TUR). 
We summarize our results and conclude in Sec. \ref{sec:summary}.

\section{Unified quantum master equation}
\label{sec:UQME}

The derivation of the Unified quantum master equation is found in Ref. \cite{trushechkin2021} 
and summarized here for completeness. 
An open system interacting with heat baths is generally described by a Hamiltonian of the form,
\be\label{eq:ham_general}
    H = H_S + H_B + V,
\ee
where $H_S$ is the Hamiltonian of the system, taken to have a discrete spectrum of eigenvalues, 
$\{E_a\}$, the differences between which comprise the set of all Bohr frequencies, $\omega_{ab}=E_a-E_b$
 (here and in what follows, we take $\hbar=k_B=1$). 
The system interacts with some number of baths labelled by $j$, each with its own free Hamiltonian $H_{B,j}$. All baths are independent of one another and do not interact directly, so $H_B = \sum_jH_{B,j}$. Finally, the interaction Hamiltonian has a contribution from each bath $j$ which can be written as a sum over direct products of operators, labelled by $\mu$, acting on the system and relevant bath Hilbert spaces:
\be
    V = \sum_j V_j = \sum_{j,\mu} S_{j\mu}\otimes B_{j\mu} = \sum_\alpha S_\alpha\otimes B_\alpha.
\label{eq:VV}
\ee
Here, $S_{j\mu}$ is an operator of the system which couples to the $j$th bath through the bath operator $B_{j\mu}$;  
 $j$ labels the bath and $\mu$ labels the (potentially many) contributions to the interaction with an individual bath.
In the last member of the equality (\ref{eq:VV}), the index $\alpha$ represents the pair of indices $j\mu$. 

As is common in quantum master equation derivations, the initial state of the total system plus baths is taken to 
be a direct product $\rho_0\otimes\prod_j \bar{\rho}_j$, where $\rho_0$ is the initial reduced density operator of the system, and $\bar{\rho}_j$ is a stationary state with respect to the bath Hamiltonian $H_{B,j}$. 
We will take $\bar{\rho}_j$ to be a Gibbs' state characterized by the temperature $T_j$ of the associated bath.

The total density operator is understood to evolve unitarily with respect to the full Hamiltonian $H$. 
The reduced system density operator is then obtained by tracing over all bath degrees of freedom,
\be 
    \rho(t) = \textrm{Tr}_B[\rho_{tot}(t)] = \textrm{Tr}_B\big[e^{-iHt}(\rho_0\otimes\prod_j\bar{\rho}_j)e^{iHt}\big].
\ee
The standard approach towards obtaining a time-local equation for the time derivative of $\rho(t)$ consists of making the 
Born and Markov approximations \cite{breuer-book}. 
The former states that the system-bath state remains a product state at all times up to second order in the interaction strength. The latter is the approximation that the baths are ``memory-less", on account of the fact that the bath correlation timescale is significantly faster than the timescale for the dynamics of interest. Thus, knowledge of the current state of the system is sufficient to determine its time derivative. The resulting equation is the Redfield quantum master equation.
 In the interaction (I) picture \cite{breuer-book,farina2019},
\begin{align}\label{eq:redfield}
    \frac{d}{dt}\rho_I(t) =& -i[H_{LS},\rho_I(t)]+\sum_{\omega,\omega'}\sum_{\alpha,\beta}\gamma_{\alpha\beta}(\omega,\omega')
    \nonumber\\
    \times& e^{i(\omega'-\omega)t}\bigg(S_{\beta\omega}\rho_I(t)S^\dag_{\alpha\omega'} - \frac{1}{2}\{S^\dag_{\alpha\omega'}S_{\beta\omega},\rho_I(t)\}\bigg).
\end{align}
We sum over all pairs of Bohr frequencies and consider ``jump" operators associated with pairs of states ($P_i$ is the projector onto eigenstate $|i\ra$),
\be
    S_{\alpha\omega} = \sum_{i,i'; \omega_{ii'}=\omega}P_{i'}S_\alpha P_i
\ee
All information about the baths is contained in the Lamb Shift Hamiltonian, $H_{LS}$,  and the rates, $\gamma_{\alpha\beta}(\omega,\omega')$ \cite{breuer-book}, which are each given in terms of Fourier transforms of bath correlation functions,
\bea\label{eq:gamma_breakdown}
    H_{LS} &=& \sum_{\omega,\omega'}\sum_{\alpha,\beta}e^{i(\omega-\omega')t}\sigma_{\alpha\beta}(\omega,\omega')S^\dag_{\alpha\omega'}S_{\beta\omega}
    \nonumber\\    
    \gamma_{\alpha\beta}(\omega,\omega') &=& \Gamma_{\alpha\beta}(\omega) + \Gamma^*_{\beta\alpha}(\omega')
    \nonumber\\
    \sigma_{\alpha\beta}(\omega,\omega') &=& \frac{1}{2i}[\Gamma_{\alpha\beta}(\omega) - \Gamma^*_{\beta\alpha}(\omega')]
    \nonumber\\
    \Gamma_{\alpha\beta}(\omega) &=& \int_0^\infty d\tau e^{i\omega\tau}C_{\alpha\beta}(\tau)
    \nonumber\\
    C_{\alpha\beta}(\tau) &=& \textrm{Tr}[e^{-iH_B\tau}B^\dag_\alpha e^{iH_B\tau}B_\beta\prod_j\bar{\rho}_j].
\eea
In general, differing values of $\omega$ and $\omega'$ allow the matrix $\gamma_{\alpha\beta}(\omega,\omega')$ 
not to be positive semidefinite. Thus, the Redfield equation is not of GKLS form and may not preserve positivity 
of the reduced density operator \cite{trushechkin2021,hartmann2020,archak2022}. 
The full secular approximation resolves this issue by neglecting all terms in the summation of
 Eq.~(\ref{eq:redfield}) for which $\omega\neq\omega'$. 
It is commonly argued that the oscillatory factor in these terms render their contirbution to the dynamics negligible 
over the timescales of interest \cite{breuer-book,manzano2020}.

\begin{figure}
    \centering
    \includegraphics[width=0.95\columnwidth, trim=45 100 45 100]{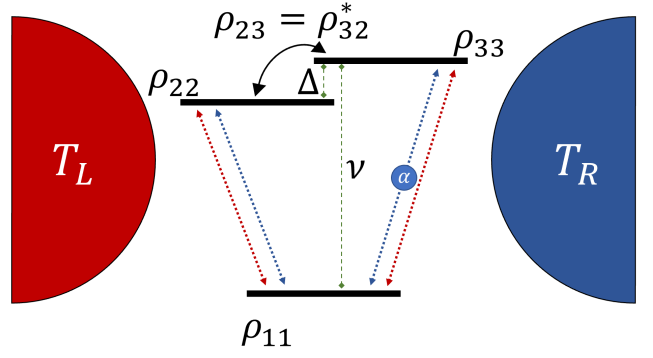}
    \caption{Schematic diagram of the ``V Model": 
a sole ground state couples to two nearly degenerate excited states. 
In the situation depicted, this coupling is realized via interactions with a hot ($L$) and cold ($R$) bath. 
The splitting $\nu$ separates between levels $|1\ra$ and $|3\ra$; levels $|2\ra$ and $|3\ra$ are splitted by energy $\Delta\ll\nu$. The $|1\ra\leftrightarrow|3\ra$ transition induced by the $R$ bath is scaled by a real-valued factor of $\alpha$ in the interaction Hamiltonian. 
All matrix elements relevant to the dynamics of the reduced density matrix  $\rho$
are depicted with their associated energy levels (or pairs of energy levels). 
The off-diagonal elements depicted 
couple to the diagonal elements and generally  
do not vanish at steady state under evolution by the Unified quantum master equation.}
\label{fig:V-diagram}
\end{figure}

However, this argument {\it fails} for systems with nearly degenerate energy levels \cite{trushechkin2021,hartmann2020}. 
More precisely, oscillatory terms cannot be neglected when they involve pairs of Bohr frequencies whose separation 
rivals the system-bath interaction energy scale in magnitude.
This includes the so-called ``V" system depicted in Fig. \ref{fig:V-diagram}: 
two excited states, both separated from the ground state approximately by energy $\nu$, are separated from each other 
only by the small splitting $\Delta\ll\nu$. 
For this system, eigenstate coherences between the two excited states, 
quantified by the density matrix elements $\rho_{23}(t)$ and $\rho_{32}(t)$, 
are expected to play a significant role in the overall dynamics
once $\Delta \leq 
\gamma(\nu)$, with $\gamma(\nu)$ 
defined above Eq. (\ref{eq:uqme_int}). 

Thus, to avoid the pitfalls of the Redfield equation without making the full secular approximation, 
we turn to the Unified quantum master equation. 
Its derivation begins with the identification of ``clusters" of 
nearly degenerate states of the system. 
The system Hamiltonian may then be represented as
\be 
    H_S = H_S^{(0)} + \delta H_S.
\ee
Clusters of nearly degenerate levels appear in $H_S^{(0)}$ as exactly degenerate, 
while $\delta H_S$ expresses only the small differences in energies within each cluster. 
A smaller set of Bohr frequencies, $\bar{\omega}$, can be identified for $H_S^{(0)}$, with each value of $\bar{\omega}$ representing a cluster, $\mathcal{F}_{\bar{\omega}}$, of nearly degenerate levels in the true system. The UQME is obtained from the Redfield 
equation by making the secular approximation {\it only with respect to well-separated levels}. That is, terms in Eq.~(\ref{eq:redfield}) with oscillatory factors $e^{i(\omega'-\omega)t}$ are neglected
only if $\omega$ and $\omega'$ fall in different clusters so $\omega'-\omega\approx\bar{\omega}'-\bar{\omega}\neq0$. 
These terms do oscillate on a faster timescale than that of the dissipative dynamics. 
Oscillating terms associated with pairs of nearby Bohr frequencies remain in the equation, 
but the rates are evaluated at these frequencies' common cluster centre 
to ensure complete positivity: 
$\gamma_{\alpha\beta}(\omega,\omega')\approx\gamma_{\alpha\beta}(\bar{\omega},\bar{\omega})\equiv\gamma_{\alpha\beta}(\bar{\omega})$. 
These approximations yield the UQME \cite{trushechkin2021},
\begin{align}\label{eq:uqme_int}
    \frac{d}{dt}\rho_I(t) = -&i[H_{LS},\rho_I(t)]
    \nonumber\\
    +&\sum_{\bar{\omega}}\sum_{\omega,\omega'\in\mathcal{F}_{\bar{\omega}}}\sum_{\alpha,\beta}e^{i(\omega'-\omega)t}\gamma_{\alpha\beta}(\bar{\omega})
    \nonumber\\
    &\times\bigg(S_{\beta\omega}\rho_I(t)S_{\alpha\omega'}^\dag - \frac{1}{2}\{S_{\alpha\omega'}^\dag S_{\beta\omega},\rho_I(t)\}\bigg).
\end{align}
In the Schrodinger picture, the explicit phase factor disappears,
 and the summation over phase factors may be incorporated into the definitions of 
operators $S_{\alpha\bar{\omega}} = \sum_{\omega\in\mathcal{F}_{\bar{\omega}}}S_{\alpha\omega}$, leading to
\cite{trushechkin2021}
\begin{align}
\label{eq:uqme}
    &\frac{d}{dt}\rho(t) = -i[H_S + H_{LS},\rho(t)]
    \nonumber\\
    &+\sum_{\bar{\omega}}\sum_{\alpha,\beta}\gamma_{\alpha\beta}(\bar{\omega})
    \bigg(S_{\beta\bar{\omega}}\rho(t)S_{\alpha\bar{\omega}}^\dag - \frac{1}{2}\{S_{\alpha\bar{\omega}}^\dag S_{\beta\bar{\omega}},\rho(t)\}\bigg).
\end{align}
Unlike the fully Secular master equation, 
this equation couples the time evolution of level populations with that of coherences 
between pairs of nearly degenerate eigenstates. 
However, it is of GKLS form on account of the positive semidefinite nature of the matrix 
$\gamma_{\alpha\beta}(\bar{\omega})$, and thus preserves the positivity of the reduced density operator $\rho$\cite{trushechkin2021}.

\section{Full counting statistics and fluctuation symmetry}\label{sec:FCS}

\subsection{Counting field-dependent UQME}

We wish to assess the validity of the fluctuation theorem for heat exchange with thermal baths under open system evolution 
governed by the UQME. 
As these theorems serve as the microscopic basis for the Second Law of Thermodynamics 
at the level of averages, this is central to the investigation of the thermodynamic consistency of this quantum master equation. 
The fluctuation theorem is closely-linked to symmetries of the scaled cumulant generating function (CGF), 
$\mathcal{G}(\boldsymbol{\chi})$, which we may study by introducing a counting field vector, 
$\boldsymbol\chi$, to the second-order master equation \cite{esposito2009,hava2018}.

In particular, we are interested in the statistics of the energy current, $J_j$, flowing from each bath $j$, which can be understood in terms of the energy change of the bath itself, $H_{B,j}(0) - H_{B,j}(t) = \int_0^t J_j(\tau)d\tau$. $J_j$ is a stochastic variable due to the Hamiltonian's status as a quantum of observable, and we can define a moment generating function for these energy changes as
\be\label{eq:mgf}
    \mathcal{Z}(\bs\chi,t) \equiv \textrm{Tr}\big[ e^{i\sum_j\chi_jH_{B,j}(0)}e^{-i\sum_j\chi_jH_{B,j}(t)} \rho_{tot}(0)\big],
\ee
where the counting field vector $\bs\chi$ is composed of elements $\chi_j$ associated with individual baths. The derivatives of the function with respect to $\chi_j$ evaluated at $\bs\chi=0$ give the moments of the distribution for the energy change in bath $j$, valid for all times. The scaled CGF, whose derivatives give the cumulants of the thermal energy current at long times, is then given by \cite{esposito2009}, 
\be\label{eq:cgf_def}
    \mathcal{G}(\bs\chi) = \lim_{t\rightarrow\infty}\frac{1}{t}\ln\mathcal{Z}(\bs\chi,t).
\ee
We will show how Eqs.~(\ref{eq:mgf}) and (\ref{eq:cgf_def}) can be derived from a modified (so-called {\it tilted}) version of the UQME.

The moment generating function $\mathcal{Z}(\bs\chi,t)$ can be understood to be the trace 
over a counting field-dependent analogue of the total density operator, $\rho_{tot}^\chi(t)$, 
which is given by the initial density operator, time-evolved by a modified propagator, 
$U^{-\chi}(t)$. This propagator is, in turn, given by the solution to the Schr\"{o}dinger equation 
if a $\bs\chi$-dressed Hamiltonian is used: $U^{-\chi}(t) = \exp[-iH^{-\chi}t]$, with
\bea\label{eq:H_chi}
    H^{-\chi} &=& e^{-i\sum_j\frac{\chi_j}{2}H_{B,j}}He^{i\sum_{j'}\frac{\chi_{j'}}{2}H_{B,j'}}
    \nonumber\\
    &=& H_S + H_B + V^{-\chi}.
\eea
As a consequence of commutation relations, the introduction of counting fields only impacts the interaction terms, which take the form,
\be 
    V^{-\chi} = \sum_j e^{-i\frac{\chi_j}{2}H_{B,j}} V_j e^{i\frac{\chi_j}{2}H_{B,j}}.
\ee
The counting field-dressed total density operator evolves according to an equation analogous to the Liouville-von Neumann equation. In the interaction picture,
\be
    \frac{d}{dt}\rho_{tot,I}^\chi(t) = -iV_I^{-\chi}(t)\rho_{tot,I}^\chi(t) + i\rho_{tot,I}^\chi(t)V_I^{\chi}(t).
\ee
In the weak coupling regime, we may derive an equation of motion for the counting field-dressed system reduced density operator $\rho^{\chi} = \textrm{Tr}_B[\rho_{tot}^\chi]$ by making the same approximations as in Sec. \ref{sec:UQME} but substituting Eq.~(\ref{eq:H_chi}) as the Hamiltonian. We arrive at the counting field-dressed Unified quantum master equation. Breaking the equation down into unitary and dissipative contributions,
\begin{align}\label{eq:uqme_chi}
    \frac{d}{dt}\rho^{\chi}(t) &= \mathcal{L}(\bs\chi)\rho^\chi(t)
    \nonumber\\
    &= -i[H_S + H_{LS},\rho^\chi(t)] + \sum_j\mathcal{D}^{j,\chi}[\rho^\chi(t)].
\end{align}
We note that the counting field does not arise in the Lamb Shift Hamiltonian. Each bath has an associated $\bs\chi$-dressed dissipator,
\begin{align} 
    \mathcal{D}^{j,\chi}[\rho^\chi(t)] = \sum_{\bar{\omega}}\sum_{\mu,\nu} &\gamma^\chi_{j\mu,j\nu}(\bar{\omega})S_{j\mu\bar{\omega}}\rho^\chi(t) S_{j\nu\bar{\omega}}^\dag
    \nonumber\\
    -&\frac{1}{2}\gamma_{j\mu,j\nu}(\bar{\omega})\{S_{j\nu\bar{\omega}}^\dag S_{j\mu\bar{\omega}},\rho^\chi(t)\}.
\end{align}
The counting field only impacts the first of the two contributions, 
as shown here, translating the argument of the correlation function by $-\chi_j$ \cite{hava2018} (provided $[H_{B,j},\bar{\rho}_j]=0$, as we do assume):
\bea 
    \gamma^\chi_{j\mu,j\nu}(\bar{\omega}) &=& \Gamma^\chi_{j\mu,j\nu}(\bar{\omega}) + \Gamma^{\chi,*}_{j\nu,j\mu}(\bar{\omega})
    \nonumber\\
    \Gamma^\chi_{j\mu,j\nu}(\bar{\omega}) &=& \int_0^\infty d\tau e^{i\bar{\omega}\tau}C^\chi_{j\mu,j\nu}(\tau)
    \nonumber\\
    C^\chi_{j\mu,j\nu}(\tau) &=& C_{j\mu,j\nu}(\tau-\chi_j).
\eea

\subsection{Proof of fluctuation symmetry}

The moment generating function is formally given by
\be\label{eq:mgf_fs} 
    \mathcal{Z}(\bs\chi,t) = \textrm{Tr}[e^{\mathcal{L}(\bs\chi)t}\rho_0],
\ee
thus the energy transport statistics are determined by the eigenvalues $\lambda(\bs\chi)$ of $\mathcal{L}(\bs\chi)$. The symmetries of these eigenvalues carry through to the CGF. The eigenvalues 
are the solutions of the characteristic equation,
\be 
    C(\bs\chi,\lambda) = \det(\mathcal{L}(\bs\chi) - \lambda(\bs\chi) I) = 0.
\ee
Here $I$ is the identity matrix.
The remainder of this section is devoted to proving the fluctuation symmetry of the moment generating function,
\be\label{eq:Z_sym}
    \mathcal{Z}(\bs\chi,t) = \mathcal{Z}(-\bs\chi - i\bs\beta,t),
\ee
where $\bs\beta$ is a vector whose elements are the inverse temperatures of the baths, $\beta_j = 1/(k_BT_j)$. We will do so by examining the properties of the characteristic function, $C(\bs\chi,\lambda)$, as well as the counting field-dependent generator of time translations, $\mathcal{L}(\bs\chi)$, itself.

We assume that the interaction Hamiltonian, when written as a sum over terms of the form $S_\alpha\otimes B_\alpha$, consists only of system coupling operators with real-valued matrix elements. This includes typical raising and lowering operators of the form $|a\ra\la b|$, but does not account for more exotic interactions in which phase shifts are associated with transitions in the system.

For any $N$-level quantum system interacting with thermal baths, it is possible to represent the counting field-dependent reduced density operator in superoperator notation, as an $N^2$-dimensional column vector with $N$ elements (say, the first $N$) representing $\chi$-dressed analogues of level populations, and the remaining $N(N-1)$ elements corresponding to the $\chi$-dressed analogues of coherences. Then $\mathcal{L}(\bs\chi)$ is an $N^2\times N^2$ matrix which acts on the vector $\rho^{\chi}(t)$ to get its time derivative. However, there is no general prescription determining how to write $\mathcal{L}(\bs\chi)$ in superoperator notation; it depends on the specifics of the system being considered, namely, which coherences are relevant to dynamics under the UQME.

We wish to show that this matrix satisfies the following property \cite{schaller2016, esposito2022}:
\be
 \mathcal{L}^\dag(-\bs\chi-i\bs\beta) = \mathcal{L}^R(\bs\chi),
\ee
where $\mathcal{L}^R(\bs\chi)$ is the generator of time translations for the time-reversed protocol. Since we consider here a Hamiltonian with no explicit time dependence, the time-reversed generator is given simply by $\mathcal{L}^R(\bs\chi) = \mathcal{L}^*(\bs\chi)$\cite{esposito2022}. Thus, it suffices to show that \cite{gernot-book}
\be\label{eq: L_sym}
    \mathcal{L}^{T}(-\bs\chi - i\bs\beta) = \mathcal{L}(\bs\chi).
\ee
To do so, it will be useful to understand how individual elements of $\rho^{\chi}(t)$ in the energy eigenbasis evolve under Eq.~(\ref{eq:uqme_chi}). Firstly, we note that the unitary part of $\mathcal{L}(\bs\chi)$ is $\chi$-independent as it refers only to the system Hamiltonian $H_S$ and the Lamb Shift Hamiltonian $H_{LS}$. Furthermore, this contribution to the generator manifests only in the diagonal elements of $\mathcal{L}(\bs\chi)$ (in superoperator notation), as
\bea\label{eq:unitary_part}
    &&-i\la a|[H_S+H_{LS},\rho^\chi(t)]|b\ra 
    \nonumber\\
    &=& -i(H_{S,aa}+H_{LS,aa}-H_{S,bb}-H_{LS,bb})\rho_{ab}^\chi(t)
    \nonumber\\
    &=& -i(\omega_{ab})\rho^\chi_{ab}(t),
\eea
where $\omega_{ab}$ has been redefined to incorporate the impacts of the Lamb Shift. Note that the unitary evolution of an element of $\rho^\chi_{ab}(t)$ depends on no elements of $\rho^\chi(t)$ besides itself. Importantly, diagonal and off-diagonal elements with respect to the original $N$-dimensional Hilbert space should not be confused with those of $\mathcal{L}(\bs\chi)$ in superoperator notation--Eq.~(\ref{eq:unitary_part}) holds even if $a\neq b$ and still describes diagonal elements of $\mathcal{L}(\bs\chi)$ in superoperator notation (the dependence of an element of $\rho^\chi(t)$ on itself). 
Therefore, the unitary part of $\mathcal{L}(\bs\chi)$ satisfies Eq.~(\ref{eq: L_sym}) trivially.

We now consider the dissipative contribution to the time evolution of $\rho^\chi(t)$. It suffices to show that the $\bs\chi$-dependent dissipator associated with any one bath $j$ satisfies Eq.~(\ref{eq: L_sym}). It is useful to write explicitly the dissipative contribution of bath $j$ towards the time derivative of a matrix element $\rho^\chi_{ab}(t)$,
\begin{widetext}
\be\label{eq:chi_uqme_elements}
    \la a|\mathcal{D}^{j,\chi}[\rho^\chi(t)]|b\ra = 
    \sum_{\bar{\omega}}\bigg[ \sum_{c,d} \gamma^{j,\chi}_{(ac),(bd)}(\bar{\omega})\rho^\chi_{cd}(t)
    -\frac{1}{2}\sum_{c,d}\gamma^{j}_{(dc),(da)}(\bar{\omega})\rho^\chi_{cb}(t) -\frac{1}{2}\sum_{c,d}\gamma^{j}_{(db),(dc)}(\bar{\omega})\rho^\chi_{ac}(t)\bigg],
\ee
\end{widetext}
where the bath-specific rates associated with given transitions are
\be\label{eq:chi_rates_elements}
    \gamma^{j,\chi}_{(ab),(cd)}(\bar \omega) = \delta_{\bar\omega,\omega^{(0)}_{ba}}\delta_{\bar\omega,\omega^{(0)}_{dc}}\sum_{\mu,\nu}\gamma^\chi_{j\mu,j\nu}(\bar\omega)\la a|S_{j\nu}|b\ra\la d|S^\dag_{j\mu}|c\ra.
\ee
We note that the counting field-dependence of these rates manifests as a complex phase, $\gamma_{(ab),(cd)}^{j,\chi}(\bar\omega) = e^{-i\omega_{ba}^{(0)}\chi_j}\gamma^j_{(ab),(cd)}(\bar\omega)$, and that the rate is nonzero only when evaluated at a Bohr frequency, $\bar{\omega}$, characterizing a cluster of levels of the system.

We want to identify the different ways that elements of $\rho^\chi(t)$ can depend on one another. This includes transfer of ($\bs\chi$-dependent) population from one level to another. However, since the master equation is nonsecular, it also includes transfer from population to coherence and coherence to population, as well as the dependence of coherences on other coherences. In order to study each of these dependencies in detail, it is useful to re-write Eq.~(\ref{eq:chi_uqme_elements}) for the specific cases of diagonal and off-diagonal elements of $\rho^\chi(t)$:
\begin{widetext}
\begin{align}\label{eq:pop_evolution}
    \la a|\mathcal{D}^{j,\chi}[\rho^\chi(t)]|a\ra &= \sum_{\bar{\omega}}\bigg[\sum_{c}e^{-i\bar{\omega}\chi_j}\gamma^j_{(ac),(ac)}(\bar{\omega})\rho_{cc}^{\chi}(t) + \sum_{c,d;c\neq d}e^{-i\bar{\omega}\chi_j}\gamma^j_{(ac),(ad)}(\bar{\omega})\rho_{cd}^{\chi}(t)
    \nonumber\\
    &- \sum_{c}\gamma_{(ca),(ca)}^j(\bar{\omega})\rho^\chi_{aa}(t) - \frac{1}{2}\sum_{c,d;c\neq a}\gamma^j_{(dc),(da)}(\bar{\omega})\rho^\chi_{ca}(t) - \frac{1}{2}\sum_{c,d;c\neq a}\gamma^j_{(da),(dc)}(\bar{\omega})\rho^\chi_{ac}(t)\bigg]
    \\
    \la a|\mathcal{D}^{j,\chi}[\rho^\chi(t)]|b\ra &= \sum_{\bar{\omega}}\bigg[\sum_c e^{-i\bar{\omega}\chi_j}\gamma^j_{(ac),(bc)}(\bar{\omega})\rho_{cc}^{\chi}(t) + \sum_{c,d;c\neq d}e^{-i\bar{\omega}\chi_j}\gamma^j_{(ac),(bd)}(\bar{\omega})\rho_{cd}^{\chi}(t)\label{eq:coh_evolution}
    \nonumber\\
    -\sum_{d;d\neq a,d\neq b} &\gamma^j_{(db),(da)}(\bar{\omega})\bigg(\frac{\rho_{aa}^\chi(t) + \rho_{bb}^\chi(t)}{2} \bigg) - \frac{1}{2}\sum_{c,d;c\neq b,d\neq a}\gamma^j_{(dc),(da)}(\bar{\omega})\rho^\chi_{cb}(t) - \frac{1}{2}\sum_{c,d;c\neq a, d\neq b}\gamma^j_{(db),(dc)}(\bar{\omega})\rho_{ac}^\chi(t)\bigg].
\end{align}
\end{widetext}

In Eq.~(\ref{eq:coh_evolution}) we assume $a\neq b$. Inspecting Eqs.~(\ref{eq:pop_evolution}) and (\ref{eq:coh_evolution}), we can identify classes of matrix elements of $\mathcal{D}^{j,\chi}$ if it were to be written in superoperator notation: those which couple diagonal elements of $\rho^\chi(t)$ to other diagonal elements, those which couple diagonal elements to off-diagonals, etc. This allows us to identify the explicit forms of pairs of matrix elements that are swapped under transposition. It is these pairs that we hope to prove are related by Eq.~(\ref{eq: L_sym}).

For example, an off-diagonal element $\rho_{cd}^\chi(t)$ may contribute via $\mathcal{D}^{j,\chi}$ to $\dot{\rho}_{aa}^\chi(t)$, provided there exists a cluster of Bohr frequencies characterized by the value $\bar{\omega}$ such that $\omega_{ca}^{(0)}=\omega_{da}^{(0)}=\bar{\omega}$. The coefficient associated with this dependence is given by the second term of Eq.~(\ref{eq:pop_evolution}),

\begin{align}
    &\sum_{\bar{\omega}}e^{-i\bar{\omega}\chi_j}\gamma^j_{(ac),(ad)}(\bar{\omega})\nonumber\\
    =&\sum_{\bar{\omega}}e^{-i\bar{\omega}\chi_j}\delta_{\bar{\omega},\omega_{ca}^{(0)}}\delta_{\bar{\omega},\omega_{da}^{(0)}}\sum_{\mu,\nu}\gamma_{j\mu,j\nu}(\bar{\omega})\la a|S_{j\nu}|c\ra\la d|S^\dag_{j\mu}|a\ra.
\end{align}
%
The corresponding coefficient is that which characterizes the dependence of $\dot \rho^\chi_{cd}(t)$ on $\rho^\chi_{aa}(t)$, which is given by the first term under the summation in Eq.~(\ref{eq:coh_evolution}) with appropriate relabelling of indices,
\begin{align}
    &\sum_{\bar{\omega}}e^{-i\bar{\omega}\chi_j}\gamma^j_{(ca),(da)}(\bar{\omega})
    \nonumber\\
    =&\sum_{\bar{\omega}}e^{-i\bar{\omega}\chi_j}\delta_{\bar{\omega},\omega^{(0)}_{ac}}\delta_{\bar{\omega},\omega^{(0)}_{ad}}\sum_{\mu,\nu}\gamma_{j\mu,j\nu}(\bar{\omega})\la c|S_{j\nu}|a\ra\la a|S^\dag_{j\mu}|d\ra
\end{align}

Making the substitution $\chi_j\rightarrow -\chi_j-i\beta_j$ here, we get,
\bea 
    &&\sum_{\bar{\omega}}e^{i\bar{\omega}\chi_j}e^{-\beta_j\bar{\omega}}\gamma^j_{(ca),(da)}(\bar{\omega})
    \nonumber\\
    =&& \sum_{\bar{\omega}}e^{-i\bar{\omega}\chi_j}e^{\beta_j\bar{\omega}}\gamma^j_{(ca),(da)}(-\bar{\omega})
    \nonumber\\
    =&& \sum_{\bar{\omega}}e^{-i\bar{\omega}\chi_j}\gamma^j_{(ac),(ad)}(\bar{\omega}),
\eea
where in the first step, we utilize the redefinition $\bar{\omega}\rightarrow-\bar{\omega}$. This is permitted as, by construction, for each cluster of Bohr frequencies centred around $\bar{\omega}$, there is a corresponding cluster at $-\bar{\omega}$ of all the same transitions in the reverse direction.

\begin{figure*}
    \centering
    \includegraphics[width=0.95\textwidth, trim=75 10 75 20]{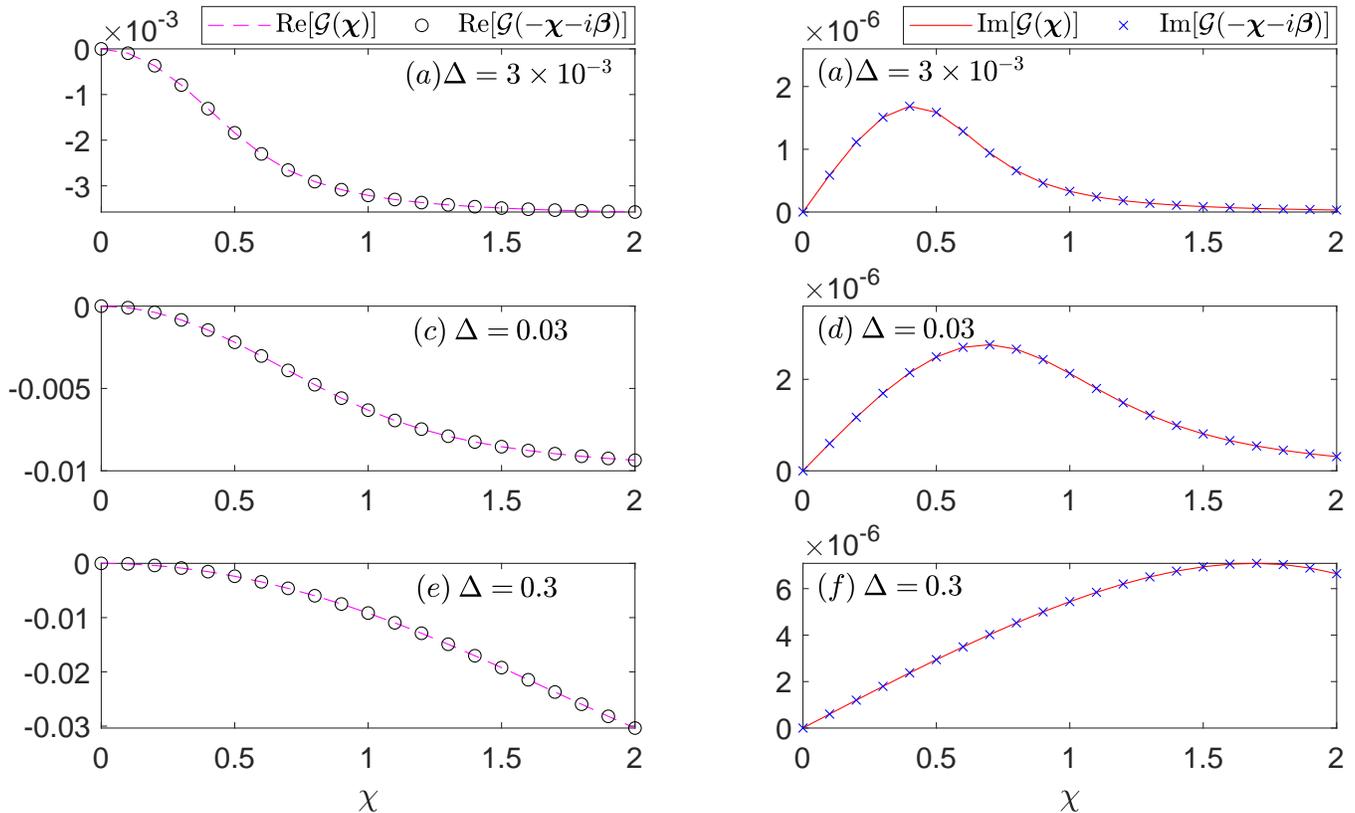}
    \caption{The real and imaginary parts of the cumulant generating function at steady state for the V system, when calculated using the Unified QME, evaluated both at $\bs\chi =(\chi, 0)^T$
    (lines) and the shifted counting field $-\bs\chi-i\bs\beta$ (markers);
the notation $\bs\chi =(\chi, 0)^T$ indicates that counting is performed on the $L$ bath only.
    The coincidence of the two curves at a range of values of $\Delta$ indicates that the CGF exhibits fluctuation symmetry. We used $T_L=4$, $T_R=3.99$, $\nu=1$, $\alpha=0.5$, and $a=0.01$; for details over the model, see Sec. \ref{sec:V_model}.
    }
    \label{fig:V-CGF-unified}
\end{figure*}

In the second step we use the assumption that matrix elements of the system coupling operators $S_{j\mu}$ are real. Each term in one summation then appears in the other. For example, suppose one of the system operators is a jump operator $\sigma=|a\ra\la c|$. Then, in order to ensure Hermiticity of the Hamiltonian, the corresponding reverse jump operator $\sigma^\dag=|c\ra\la a|$ must appear as another system coupling operator. The term in the summation for one rate containing $\la a|\sigma|c\ra$ corresponds to the term in the summation for the other rate containing $\la c|\sigma^\dag|a\ra$, since $\la c|\sigma^\dag|a\ra = \la a|\sigma|c\ra^*= \la a|\sigma|c\ra$. We also use the local detailed balance relations satisfied by the rates given in Eq.~(\ref{eq:chi_rates_elements}), by virtue of the symmetry properties of the correlation functions, Eq.~(\ref{eq:gamma_breakdown}). In doing so, we have shown that this pair of matrix elements is consistent with the relation of Eq.~(\ref{eq: L_sym}).

\begin{figure*}
    \centering
    \includegraphics[width=0.95\textwidth, trim=75 10 75 20]{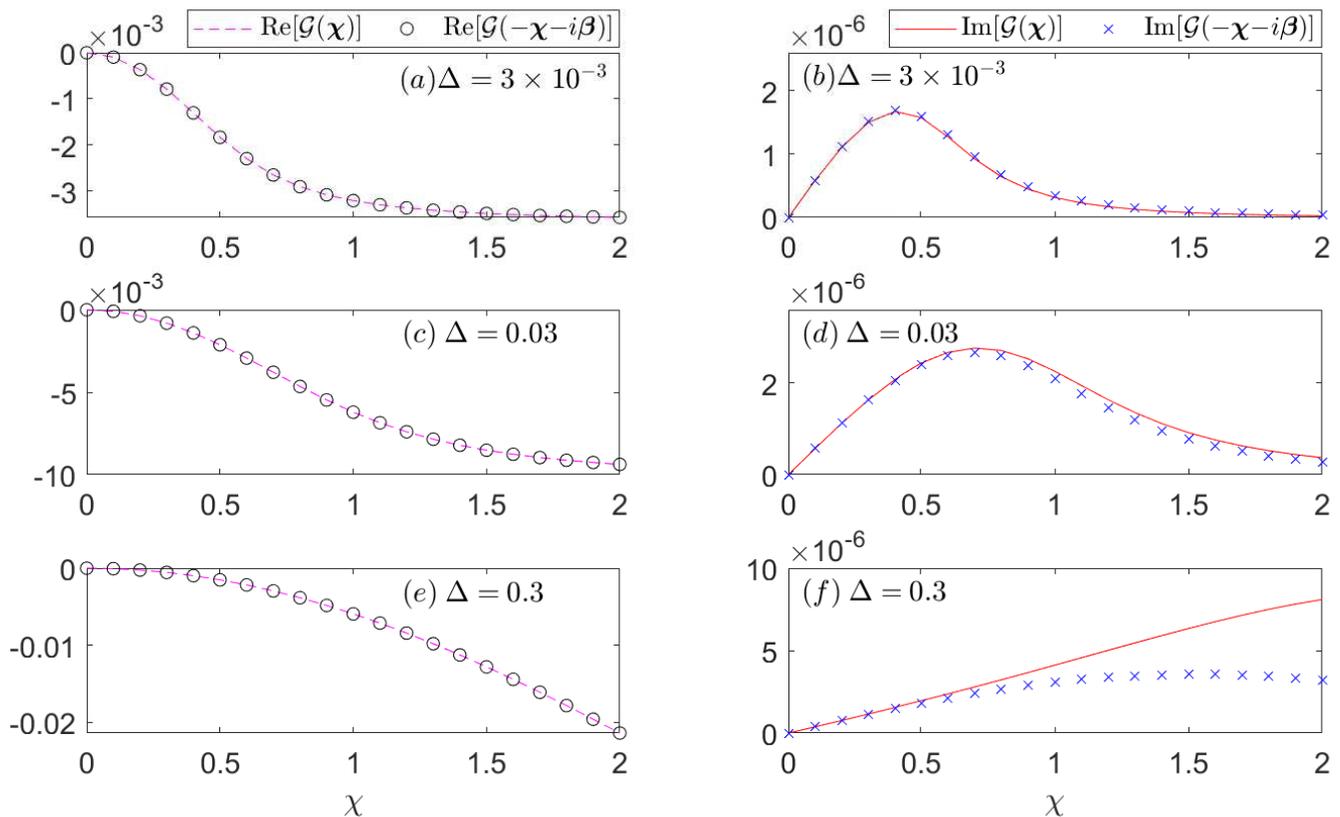}
    \caption{The cumulant generating function at steady state for the V system, when calculated using the full Redfield QME, evaluated both at $\bs\chi$ (lines) and the shifted counting field $-\bs\chi-i\bs\beta$ (markers). The violation of fluctuation symmetry is exemplified by the deviation between the two curves representing the imaginary part as $\chi$ grows. 
    Parameter values are the same as in Fig. \ref{fig:V-CGF-unified}.}
    \label{fig:V-CGF-redfield}
\end{figure*}

There are eight other classes of pairs of matrix elements which can be shown using analogous arguments to satisfy the symmetry relation expressed in Eq.~(\ref{eq: L_sym}). They represent dependencies between elements of $\rho^\chi(t)$, some of which can be understood by inspecting Fig. \ref{fig:V-diagram}. In the V model depicted here, two thermal baths can each excite transitions between a ground state and two nearly degenerate excited states, separated by $\Delta\ll\nu$. The coherences between the two excited states, associated with the elements $\rho_{23}^\chi$ and $\rho_{32}^\chi$ of the counting-dependent density operator 
are relevant to the dynamics and couple to the populations of the levels, associated with elements $\rho_{11}^\chi$, $\rho_{22}^\chi$, and $\rho_{33}^\chi$.
    

The specific equations for the V model example are given in Sec. \ref{sec:V_model}. Particularly, upon inspection of Eq.~(\ref{eq:L_V}), one can see that each of the corresponding pairs of matrix elements satisfy the necessary relation for Eq.~(\ref{eq: L_sym}) to hold.

We now write down the characteristic equation to solve for the eigenvalues of the generator evaluated at $-\bs\chi-i\bs\beta$ \cite{gernot-book}:
\bea
    \det(\mathcal{L}(-\bs\chi-i\bs\beta) - \lambda(-\bs\chi-i\bs\beta)I)&=&0,
    \nonumber\\
    \det(\mathcal{L}^T(-\bs\chi-i\bs\beta) - \lambda(-\bs\chi-i\bs\beta)I) &=&0,
    \nonumber\\
    \det(\mathcal{L}(\bs\chi) - \lambda(-\bs\chi-i\bs\beta)I) &=&0.
\eea
We arrive at the second line by noting that transposition does not impact the determinant of a matrix, and at the third line by utilizing Eq.~(\ref{eq: L_sym}). The eigenvalues for the generator at the shifted value of the counting field, $\lambda(-\bs\chi - i\bs\beta)$, are seen to be solutions of the characteristic equation for the non-shifted version, $\mathcal{L}(\bs\chi)$, thus matching its eigenvalues. We have that each $\lambda(\bs\chi)=\lambda(-\bs\chi-i\bs\beta)$, and thus the moment generating function satisfies the corresponding symmetry, Eq.~(\ref{eq:Z_sym}) {\it at all times}. So does the scaled CGF in the long time limit:
\be\label{eq:cgf_fs}
    \mathcal{G}(\bs\chi) = \mathcal{G}(-\bs\chi-i\bs\beta).
\ee
This steady state relation is demonstrated in Fig. \ref{fig:V-CGF-unified} for the V system
. Furthermore, as we discuss further in Sec. \ref{sec:V_model}, we demonstrate in Fig. \ref{fig:V-CGF-redfield} that this symmetry is violated when the Redfield QME is used instead of the Unified QME for the same model.
 
 Eq.~(\ref{eq:cgf_fs}) can be shown to be equivalent to the familiar detailed fluctuation theorem \cite{esposito2009, hanggi11,gernot-book}, which relates the steady-state probability of a process in which some amount of energy $\Delta E_j$ is delivered to each bath $j$ (characterized by a vector $\bs\Delta\bs E$) and that of its reverse process in which these quantities of energy are drawn from the baths (represented as $-\bs\Delta\bs E$):
\be\label{eq:FT}
    \lim_{t\rightarrow\infty}\frac{P_{\bs\Delta\bs E}(t)}{P_{-\bs\Delta\bs E}(t)}=e^{\sum_j\beta_j\Delta E_j} = e^{\Delta S}.
\ee
The entropy production for this steady state process is identified as the sum over amounts of energy delivered to each bath multiplied by the bath inverse temperatures $\beta_j$. Eq.~(\ref{eq:FT}) is a sufficient condition for the Second Law of Thermodynamics at the level of averages, $\langle \Delta S\rangle\geq 0$ \cite{esposito2009}.

\section{Example: The V Model}
\label{sec:V_model}

\subsection{Counting-dependent Unified quantum master equation for the V system}

We demonstrate the implications of the fluctuation symmetry, Eq.~(\ref{eq:cgf_fs}), for the three-level V system depicted in Fig. \ref{fig:V-diagram}. The total Hamiltonian is of the form of Eq.~(\ref{eq:ham_general}) with the system Hamiltonian,
\be 
    H_S = (\nu-\Delta)|2\ra\la 2| + \nu|3\ra\la3|.
\ee
Note that the state $|1\ra$ corresponds to zero energy. The UQME is valid in the limit $\Delta\ll\nu$, in which coherence between the eigenstates $|2\ra$ and $|3\ra$ is most relevant to the overall dynamics. Two bosonic baths ($L$ and $R$) at temperatures $T_L>T_R$ couple to the system and can induce the transitions $|1\ra\leftrightarrow|2\ra$ and $|1\ra\leftrightarrow|3\ra$.

The Hamiltonian of each bath is
\be
    H_{B,j} = \sum_k \omega_{k,j}b^\dag_{k,j}b_{k,j},
\ee
where $j=L,R$ and $b_{k,j}$ is the annihilation operator for a bosonic mode $k$ of bath $j$ with energy $\omega_{k,j}$. Taking the baths to be independent and each at thermal equilibrium, the number operators satisfy $\la b^\dag_{k,j}b_{k,j}\ra = n_{j}(\omega_{k,j})$, with $n_j(\omega) = [e^{\beta_j\omega}-1]^{-1}$ the Bose-Einstein distribution at temperature $T_j$.

Each bath has an associated contribution to the interaction Hamiltonian for this minimal model which can be written as a product,
\be 
    V_j = S_j\otimes B_j;\enspace B_j=\sum_k g_{k,j}(b^\dag_{k,j} + b_{k,j}).
\ee
The real numbers $g_{k,j}$ characterize the coupling strength between the system and each bath mode. Since all the baths are independent and each has only one bath operator $B_j$, the only nonvanishing correlation functions between bath operators are autocorrelation functions, which are written with just one index for simplicity, $C_j(\tau) = \la B_j(\tau)B_j(0)\ra$.

The system operators, $S_j$ differ from one another such that there is an asymmetry in the couplings for the two transitions associated with the right bath,
\bea
    S_L &=& |1\ra\la2| + |1\ra\la3| + \textrm{h.c.}\nonumber\\
    S_R &=& |1\ra\la2| + \alpha|1\ra\la3| + \textrm{h.c.},
\eea
where $\alpha$ is a real number, so as to satisfy the condition for the above proof of fluctuation symmetry.

In deriving the UQME for this system, we identify three clusters of Bohr frequencies associated with values $\{-\nu, 0, \nu\}$ (another appropriate choice would be the average energy of the excited states, $\nu-\Delta/2$, instead of $\nu$; the distinction is negligible in the limit $\Delta\ll\nu$). The lack of dissipative coupling between levels $|2\ra$ and $|3\ra$ means no rates that factor into the dynamics are associated with the cluster at $\bar{\omega}=0$. Focusing on the positive frequency cluster $\bar{\omega}=\nu$, we calculate the half Fourier transform
\be
    \Gamma_j(\nu) = \int_0^\infty d\tau e^{i\nu\tau}C_j(\tau) = \frac{\gamma_j(\nu)}{2} + iZ_j(\nu)\approx \frac{\gamma_j(\nu)}{2}.
\ee
At sufficiently high temperatures, it is possible to demonstrate that the imaginary part of the half Fourier transform is significantly less than the smallest energy scale in the problem, $Z_j(\nu)\ll\Delta$ \cite{ivander2022}. We therefore neglect it, which amounts to neglecting the Lamb shift.  $\gamma_j(\nu)$ is the rate of transitions induced by bath $j$,
\be
    \gamma_j(\nu) = \mathcal{J}_j(\nu)[n_j(\nu) + 1],
    \label{eq:gammajnu}
\ee
where $\mathcal{J}_j(\omega) = 2\pi\sum_k g^2_{k,j}\delta(\omega-\omega_{k,j})$ is the spectral density function of bath $j$, which we take to be of Ohmic form, $\mathcal{J}_j(\omega) = a\omega$, with constant coefficient $a$ the same for both baths. Local detailed balance then fixes the rates derived from the half Fourier transform evaluated at $\bar{\omega}=-\nu$,
\be 
    \gamma_j(\nu)e^{-\beta_j\nu} = \gamma_j(-\nu) = \mathcal{J}_j(\nu)n_j(\nu).
\ee
The counting field-dependent density operator has five relevant matrix elements in the system energy eigenbasis: the three diagonal elements, $\rho_{11}^\chi$, $\rho_{22}^\chi$, and $\rho_{33}^\chi$, as well as the two off-diagonals, $\rho_{23}^\chi$ and $\rho_{32}^\chi$. For this model, the remaining off-diagonals $\rho_{12}^\chi$, $\rho_{13}^\chi$, $\rho_{21}^\chi$, and $\rho_{31}^\chi$ decouple from the others even under the full Redfield equation since $S_{j,23}=S_{j,32}=0$. However, even if the required matrix elements were nonzero, the UQME would decouple these off-diagonals from the rest of $\rho^\chi(t)$ as a consequence of the large energy separation between their associated states.

Thus, the equations of motion for $\rho^\chi(t)$ may be written in superoperator notation as a 5$\times$5 matrix acting on the vector describing the quantum state, $\rho^{\chi}(t)=(\rho_{11}^\chi(t), \rho_{22}^\chi(t), \rho_{33}^\chi(t), \rho_{23}^\chi(t), \rho_{32}^\chi(t))^T$. For simplicity, we apply counting only for energy leaving the $L$ bath, so $\bs\chi=(\chi_L,\chi_R)^T\equiv(\chi,0)^T$. To write the rates without arguments, we define $k_j\equiv\gamma_j(\nu)$ and $\Tilde{k}_j\equiv\gamma_j(-\nu)$. We have $\dot{\rho}^\chi(t) = \mathcal{L}(\bs\chi)\rho^\chi(t)$, with
\begin{widetext}
\be\label{eq:L_V}
   \mathcal{L}(\bs\chi) =
    \left( {\begin{array}{ccccc}
    -2\Tilde{k}_L - (\alpha^2+1)\Tilde{k}_R    & k_Le^{-i\chi\nu} + k_R & k_Le^{-i\chi\nu} + \alpha^2k_R & k_Le^{-i\chi\nu} + \alpha k_R& k_Le^{-i\chi\nu} + \alpha k_R  \\
    \Tilde{k}_Le^{i\chi\nu} + \Tilde{k}_R    & -k_L-k_R & 0 & -\frac{1}{2}(k_L + \alpha k_R) & -\frac{1}{2}(k_L + \alpha k_R)  \\
    \Tilde{k}_Le^{i\chi\nu} + \alpha^2\Tilde{k}_R    & 0 & -k_L-\alpha^2k_R & -\frac{1}{2}(k_L + \alpha k_R) & -\frac{1}{2}(k_L + \alpha k_R)  \\
    \Tilde{k}_Le^{i\chi\nu} + \alpha \Tilde{k}_R     & -\frac{1}{2}(k_L + \alpha k_R) & -\frac{1}{2}(k_L + \alpha k_R) & i\Delta-k_L-\frac{1}{2}(\alpha^2+1)k_R & 0  \\
    \Tilde{k}_Le^{i\chi\nu} + \alpha \Tilde{k}_R    & -\frac{1}{2}(k_L + \alpha k_R) & -\frac{1}{2}(k_L + \alpha k_R) & 0 & -i\Delta-k_L-\frac{1}{2}(\alpha^2+1)k_R
    \end{array} } \right).
\ee
\end{widetext}
By inspecting Eq.~(\ref{eq:L_V}), one can see that the counting field-dependent generator for the V system satisfies the symmetry of Eq.~(\ref{eq: L_sym}). The lower right 4$\times$4 block is $\chi$-independent, and it is symmetric as needed. The transformation $\bs\chi\rightarrow-\bs\chi-i\bs\beta$ has the effect of transposing the first row and first column, which the subsequent transposition then counteracts. To see this, note that the first row and column do depend on $\chi_R$, the exponentials $e^{\pm i(0)\nu}$ are just not shown explicitly. The transformation then entails that $\chi_R\rightarrow-i\beta_R$. Thus, factors of $e^{\pm\beta_j\nu}$  are introduced to each term, swapping the rates $k_j$ and $\Tilde{k}_j$ as needed.

It is clear from the structure of the counting field-dependent generator shown here how the arguments we employ extend to more general systems. Namely, the UQME satisfies fluctuation symmetry for any system with thermally activated transitions between clusters of nearly degenerate states. One can identify $\bs\chi$-independent matrix blocks within the generator that relate elements of $\rho^\chi$ strictly corresponding to states within the same cluster, and other blocks that have $\bs\chi$-dependence and describe transitions between clusters.

\begin{figure*}
    \centering
    \includegraphics[width = 0.9\textwidth, trim = 20 30 20 30]{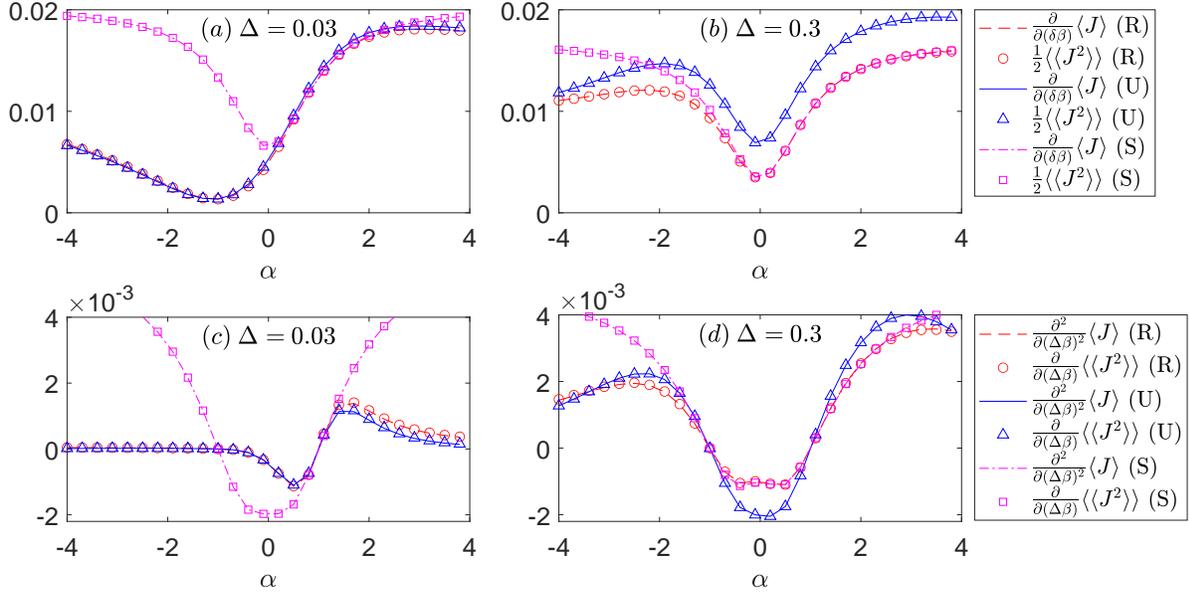}
    \caption{The V system's adherence to transport symmetries under the Redfield (red dashed, circles), Unified (blue solid, triangles) and fully Secular (magenta dash-dot, squares) master equations, calculated using full counting statistics. The coefficients are plotted here as a function of $\alpha$, the variable characterizing the coupling of the $|1\ra\leftrightarrow|3\ra$ transition to the $R$ bath. Parameter values are otherwise the same as in Fig. \ref{fig:V-CGF-unified}, amounting to the linear response regime. Upper panels: Eq.~(\ref{eq:green-kubo}), which is satisfied by all three equations, though each makes 
    different predictions for the coefficients themselves. (a) in the $\Delta = 0.03\sim \gamma$ regime, the Unified QME tracks the predictions made by the Redfield equation closely, while (b) when $\Delta=0.3\gg \gamma$, the fully Secular equation serves as a better approximation when $\alpha>0$. Lower panels: Eq.~(\ref{eq:next-sym}), for (c) $\Delta=0.03$ and (d) $\Delta =0.3$, is also generally satisfied, exhibiting similar behavior to the Green-Kubo relation.}
    \label{fig:transport}
\end{figure*}
\subsection{Numerical simulations and comparison to the full Redfield equation}

We study the heat current and its cumulants using the Redfield equation, the UQME, and the fully Secular QME.
In the fully secular case, all coherences are ignored. 
In the UQME, we cluster the energies of levels 2 and 3 into the same manifold 
and maintain their coherences, even when extending the level splitting $\Delta$ to large values (exceeding decay rates). 
This allows us to understand the regimes in which clustering should and should not be performed. 
Recall that the dissipator of the UQME is constructed from the clustered levels.
However, a proper application of the UQME should cease clustering levels 2 and 3 once  $\Delta$ exceeds the decay rates $\gamma$, 
with the UQME becoming then the standard secular treatment for the V model.

\subsubsection{Fluctuation symmetry for the V model}

The steady state scaled CGF, $\mathcal{G}(\bs\chi)$, for the V system is the eigenvalue of $\mathcal{L}(\bs\chi)$ whose real part vanishes in the limit $\bs\chi\rightarrow0$. 
Finding the full analytic expression for $\mathcal{G}(\bs\chi)$ entails solving the fifth-order characteristic polynomial, $C(\chi,\lambda)$. However, the fluctuation symmetry can be demonstrated by calculating the CGF numerically at both $\bs\chi$ and the shifted value of the counting field, $-\bs\chi-i\bs\beta$. Fig. \ref{fig:V-CGF-unified} contains plots of the CGF for the V system evaluated at both of these points for varying $\Delta$, including at $\Delta\ll \gamma(\nu)$. In this limit, the Unified quantum master equation approaches a form identical to the Redfield equation for this model.
Note that the relevant relationship when using an ohmic spectral density function [below Eq. (\ref{eq:gammajnu})], 
and at high temperature, $T>\nu$, 
is $\Delta\ll a T$, independent of frequency $\Delta$ or $\nu$. 
As $\Delta$ grows the fluctuation symmetry is still seen to be satisfied, even as we approach the regime where the approximations leading to the UQME
with levels 2 and 3 clustered 
are not valid.

When the full Redfield equation is instead used to derive the cumulant generating function,
explicit violations of fluctuation symmetry are evident in the regime that $\Delta\gtrsim a T$, and become more pronounced as $\Delta$ grows.  This is seen in Fig. \ref{fig:V-CGF-redfield}(d) and (f), as the curves corresponding to Im$[\mathcal{G}(\bs\chi)]$ and Im$[\mathcal{G}(\bs\chi-i\bs\beta)]$ diverge from one another as $\chi$ grows. Thus, detailed fluctuation theorems as given in Eq.~(\ref{eq:FT}) do not, in general, follow from Redfield descriptions of open quantum system dynamics.

We do note, however, that in our simulations, the real part of $\mathcal{G}(\bs\chi)$ for the V system derived from the Redfield QME, while taking on values slightly different from that derived from the Unifed QME, do display fluctuation symmetry (Fig. \ref{fig:V-CGF-redfield}(a), (c) and (e)). Similarly, the imaginary part of the CGF from the Redfield-based description approximately satisfies the fluctuation symmetry {\it for sufficiently small} $\chi$. 
Thus, a Redfield-based analysis involving only the first cumulant (average heat current) and second cumulant (variance) should not uncover any direct violations of the Second Law at the level of averages.

\subsubsection{Transport symmetries in linear response}

The Redfield equation can lead to violations of the fluctuation symmetry, as we demonstrated in Fig. \ref{fig:V-CGF-redfield}.
However, certain relations between transport that hold as a result of fluctuation symmetry involve only the first and second cumulants \cite{saito2008}. No violations of these relations are observed even if the Redfield equation is used.
In the linear response regime, when the temperature difference $\delta T = T_L-T_R$ is smaller than any other energy scale in the problem, this means we expect even the Redfield equation to satisfy the Green-Kubo relation,
\be\label{eq:green-kubo}
    \frac{\partial}{\partial (\delta\beta)}\langle J\rangle\big|_{\delta\beta=0} = \frac{1}{2}\langle\langle J^2\rangle\rangle\big|_{\delta\beta=0},
\ee
and the symmetry between the next-lowest order terms in the expansions for $\langle J\rangle$ and $\langle\langle J^2\rangle\rangle$,
\be\label{eq:next-sym}
    \frac{\partial^2}{\partial (\delta\beta)^2}\langle J\rangle\big|_{\delta\beta=0} = \frac{\partial}{\partial (\delta\beta)}\langle\langle J^2\rangle\rangle\big|_{\delta\beta=0},
\ee
where here, $\delta\beta=\beta_R-\beta_L$ with $\beta_j=1/T_j$ the inverse temperature, and $J\equiv J_L$, the heat current flowing from the $L$ bath. These two relations are demonstrated in Fig. \ref{fig:transport}, with, in each case, the left hand side represented by lines, and the right hand side represented by the co-incident markers.

We compare the predictions of the Redfield and Unified equations with those of the fully Secular master equation, which throws away all oscillating terms in Eq.~(\ref{eq:redfield}) rather than only those that oscillate quickly. While, as shown, each equation satisfies the transport symmetries, the value of $\Delta$ has a strong bearing on whether the Unified or fully Secular equation gives results closer to those of the less approximate Redfield equation. Generally, the $\alpha<0$ regime corresponds to the case where coherences have a more pronounced effect\cite{ivander2022}, and the Secular equation, which neglects them, tends to diverge from Redfield. 
However, when $\Delta\gg \gamma$, the UQME
with clustering of Bohr frequencies 
is invalid. 
This is because the two excited states become far apart in energy, and so it is not effective to calculate transition rates based on a {\it common} Bohr frequency $\bar{\omega}=\nu$.
Simultaneously, the timescale for the oscillations of terms coupling coherences to populations becomes sufficiently short that the full secular approximation is reasonable.
Consequently, a proper implementation of the UQME would instead recognize each of the excited states as being in a cluster of its own, returning precisely the fully Secular equation. This equation outperforms the UQME with clustering in this regime, particularly when $\alpha>0$, as seen in Fig. \ref{fig:transport}(b) and (d). 

\begin{figure}
    \centering
    \includegraphics[width = 0.875\columnwidth, trim=20 15 20 25]{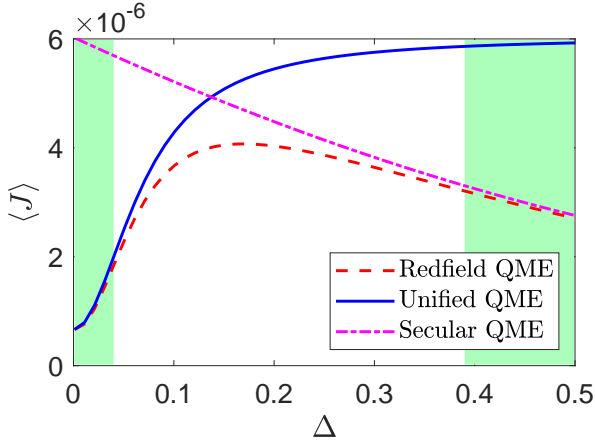}
    \caption{The mean steady-state heat current through the V system as a function of $\Delta$, given by the Redfield (red dashed), Unified (blue solid) and fully Secular (magenta dash-dot) quantum master equations. The shaded regions represent regimes in which a thermodynamically consistent GKLS master equation gives predictions similar to the Redfield equation: namely the Unified QME for small $\Delta$ and the fully Secular QME for large $\Delta$. $\alpha=-0.5$, signifying the regime in which coherences are significant; parameter values are otherwise the same as in Fig. \ref{fig:V-CGF-unified}.}
    \label{fig:crossover}
\end{figure}
\begin{figure*}[htb]
    \centering
    \includegraphics[width=0.85\textwidth, trim=30 10 30 10]{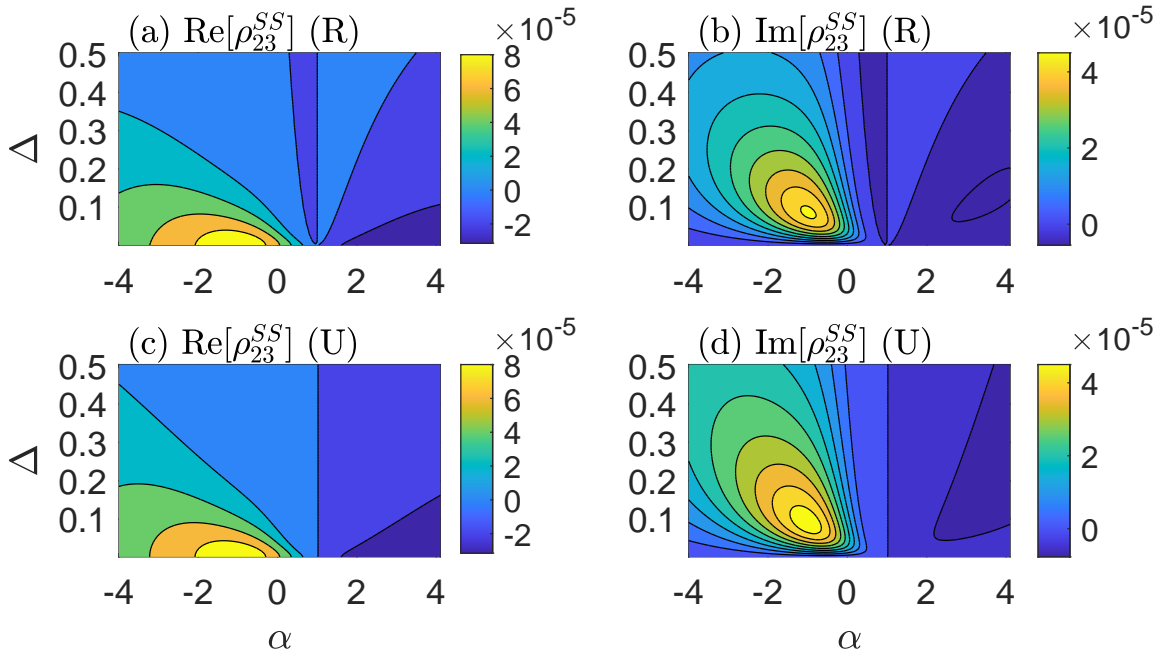}
    \caption{Steady-state values of the off-diagonal density matrix element $\rho_{23}$, as calculated using both the Redfield (upper panels) and Unified (lower panels) quantum master equations. The real (left) and imaginary (right) parts are shown as a function of $\alpha$ and $\Delta$. Parameter values are otherwise the same as in Fig. \ref{fig:transport}.}
    \label{fig:coh_contour}
\end{figure*}

\subsubsection{Validity of QMEs and the crossover region}

Focusing on the heat current,
the crossover from the regime where the Unified QME, with levels 2 and 3 clustered, is a good approximation to that where the fully Secular QME is preferred (compared to results from Redfield) can be seen more explicitly in Fig. \ref{fig:crossover}. 
Clearly, for a system with nearly-degenerate levels, the distinction between the Unified equation and the Redfield equation itself is minimal, and the Unified QME gives predictions for the mean heat current that match the Redfield equation very closely. In the opposite limit, the Secular equation's absence of coherences do not cause any significant issues, and it serves better not to cluster Bohr frequencies based on groups of levels nearby in energy. There is, however, a crossover regime (unshaded region) at which $\Delta\approx 
\gamma$, and neither the Unified with clustered excited states nor the fully Secular equation gives a prediction for the mean current that matches the Redfield equation closely. As such, it remains an open question whether, in this regime, there is a master equation description for open quantum systems that is of a GKLS form, thermodynamically consistent, and gives accurate predictions for statistics of heat currents.

 We note  that when properly using the UQME, one should refrain from clustering the excited states once their separation becomes large. In this case, results of the UQME should turn (alas in a discontinuous manner) into the fully secular approach within the unshaded region of Fig.  \ref{fig:crossover}. 

We can similarly inspect the predictions made by both the Unified and Redfield equations for the coherences at steady state, particularly the real and imaginary parts of the density matrix element $\rho_{23}$ associated with the two excited states. Once again, as seen in Fig. \ref{fig:coh_contour}, Unified and Redfield equations give rise to values that are close when $\Delta\ll \gamma$, but begin to diverge as $\Delta$ grows. In the larger $\Delta$ region, the Redfield equation predicts smaller values in the $\alpha<0$ region, especially for the imaginary part. This is more in accordance with the fully Secular master equation which predicts that all eigenstate coherences vanish at steady state (not shown).

We point out that coherences are rather small in Fig. \ref{fig:coh_contour}. This is because of the high temperature, $T\gg \nu$, and small temperature difference,
$\delta T\ll \nu$,
employed in this example.
As was shown in Ref. \citenum{ivander2022},
steady state coherences in the V model
scale as ${\rm Re}[\rho_{23}^{SS}] \propto \left(e^{-\beta_R\nu} - e^{-\beta_L\nu}\right)$ and ${\rm Im }[\rho_{23}^{SS}] \propto  \frac{\Delta}{k}{\rm Re }[\rho_{23}^{SS}]$.


\subsubsection{Thermodynamic Uncertainty Relation}

\begin{figure}
    \centering
    \includegraphics[width=0.95\columnwidth, trim=20 10 20 10]{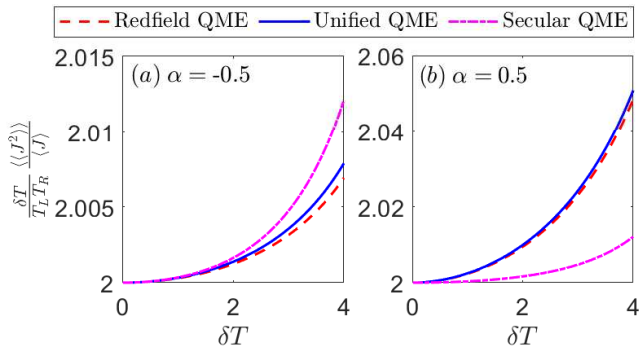}
    \caption{The TUR ratio (LHS of Eq.~(\ref{eq:TUR})) with the heat current cumulants calculated as a function of $\delta T$ using the Redfield (red dashed), Unified (blue solid), and fully Secular (magenta dash-dot) quantum master equation. $\Delta=0.03$, signifying the regime where coherences are non-negligible at steady state. The average temperature of the two baths is $\bar{T} = 4$. (a) $\alpha=-0.5$ and (b) $\alpha=0.5$. Otherwise, parameter values are the same as in Fig. \ref{fig:V-CGF-unified}.}
    \label{fig:TUR}
\end{figure}

Finally, we compare the predictions these equations make regarding the Thermodynamic Uncertainty Relation (TUR) \cite{barato2015,pietzonka2016}, a cost-precision tradeoff relation between the relative fluctuations (precision) of a steady-state current and the mean entropy production rate $\langle\sigma\rangle$. The TUR was derived originally for classical Markovian dynamics and shown to hold in certain cases for quantum heat transport \cite{thermal-TUR}. For heat transport between two reservoirs at steady state, it takes the form,
\be\label{eq:TUR} 
    \langle\sigma\rangle\frac{\langle\langle J^2\rangle\rangle}{\langle J\rangle^2} = \delta\beta\frac{\langle\langle J^2\rangle\rangle}{\langle J\rangle}\geq2.
\ee
In Fig. \ref{fig:TUR}, we plot the expression on the left hand side of this inequality, referred to as the TUR ratio, as a function of $\delta T$ (note that $\delta\beta=\delta T/T_LT_R$), with the first and second cumulants of the heat current as predicted by the Redfield, Unified and fully Secular quantum master equations. As discussed, these two cumulants of the current, even when calculated using Redfield equations, do not directly exhibit any violations of fluctuation symmetry. We see that in all cases, the TUR is satisfied as $\delta T$ increases, with the TUR ratio converging to 2 in the equilibrium limit, despite taking on different values for the three equations when $\delta T>0$. The close correspondence between the 
Unified and Redfield equations reflects the small value used for $\Delta$ in the simulations, corresponding to the case where coherences at steady state are non-negligible, as discussed above and observed also in Fig. \ref{fig:transport}.
The main nontrivial observation from Fig. \ref{fig:TUR} is that while the Secular current is significantly higher than what is predicted by the Unified or Redfield equation for the presented parameters at small $\Delta$ (recall Fig. \ref{fig:crossover}), the TUR ratio itself is almost identical between the methods,  even far from equilibrium.

\section{Summary}
\label{sec:summary}

We have investigated the question of whether the Unified quantum master equation gives rise to a cumulant generating function for the statistics of heat currents that satisfies the heat exchange fluctuation symmetry.
Our findings demonstrate that this is indeed the case for general quantum systems interacting with
thermal baths. We have therefore shown that it is possible to achieve thermodynamic consistency with a master equation-based description of open quantum systems that does not make the full secular approximation with respect to all oscillating terms in the Redfield equation. The UQME retains the effect of eigenstate coherences, which, in turn, do not vanish at steady state. Coherences are important precisely in the 
regime where the UQME differs from the fully Secular QME: 
when systems have nearly-degenerate energy levels, such that the approximations leading to the fully Secular master equation are not valid.

In particular, we have shown based on a symmetry of the 
counting field-dependent Liouvillian, $\mathcal{L}(\bs\chi)$, 
associated with the 
UQME, that {\it each of its eigenvalues} satisfies the symmetry relation $\lambda(\bs\chi)=\lambda(-\bs\chi - i\bs\beta)$. The so-called ``dominant" eigenvalue--the one whose real part approaches zero as $\bs\chi$ approaches zero--is exactly the CGF for heat current statistics at steady state. However, the satisfaction of this symmetry by {\it all} eigenvalues indicates that energy transport statistics are thermodynamically consistent at all times, not just in the long-time limit.

We have exemplified our results on the V system, a minimal model exhibiting a pair of nearly degenerate excited states, between which coherences survive in steady state transport. 
Fluctuation symmetry is manifest in the specific Unified equations of motion for the reduced density operator of this system when written out. 
We further identified  regimes in which the Redfield equation itself is approximately thermodynamically consistent, and thus leads to predictions that still satisfy transport symmetries such as the Green-Kubo relation.

Our simulations also indicate a crossover from the regime of validity of the Unified QME with the excited states clustered to that of the fully Secular QME as the energy splitting between excited states grows. Then, the impact of coherences on the overall dynamics diminishes, and the supposition that Bohr frequencies of the system can be categorized neatly into ``clusters" no longer adheres to reality. 
Thus, while the Unified QME is positivity-preserving and thermodynamically consistent regardless of which energy levels are clustered together, the spectrum of the system Hamiltonian must be taken into consideration when assessing how this clustering should be carried out, and whether a UQME should be used that differs from the fully Secular equation \cite{trushechkin2021}.
 
 The UQME allows studies of transport fluctuations while including prominent  coherences. This method can therefore be used to derive cost-precision tradeoffs in the operation of quantum thermal machines, testing results based on the Lindblad formalism \cite{Saito22}. 
Finally, the principles of the UQME can be readily applied to master equations that account for strong system-bath couplings. 
Developing e.g., a polaron-transformed \cite{PT1,PT2,cao16} UQME and a reaction-coordinate UQME \cite{RCrev,Nick1,Nick2,latune22} is left for future work.


\begin{acknowledgements}
DS acknowledges support from an NSERC Discovery Grant and the Canada Research Chair program.
The work of MG was supported by an Ontario Graduate Scholarship (OGS).
\end{acknowledgements}


\end{document}